\documentclass[twocolumn,showpacs,preprintnumbers,amsmath,amssymb]{revtex4}

\usepackage{graphicx}
\usepackage{dcolumn}
\usepackage{bm}

\begin{document}


\title{Control, measurement and propagation of entanglement in photon pairs generated through type-II parametric downconversion}

\author{Yasser Jeronimo-Moreno$^1$ and Alfred B. U'Ren$^{1,2}$}
\affiliation{$^1$Departamento de \'Optica, Centro de Investigaci\'on
Cient\'{\i}fica y de Educaci\'on Superior de
Ensenada, Apartado Postal 2732, Ensenada, BC 22860, M\'exico \\
$^2$Instituto de Ciencias Nucleares, Universidad Nacional Aut\'onoma de M\'exico, apdo. postal 70-543, M\'exico 04510 DF.}

\date{\today}

%
\newcommand{\epsfg}[2]{\centerline{\scalebox{#2}{\epsfbox{#1}}}}

\begin{abstract} We present a study of the effects of dispersion on the properties of photon pairs generated by type-II, collinear spontaneous parametric downconversion.  Specifically, we take into consideration the effects of a chirped pump, as well as of dispersive propagation of the photon pairs.   We present expressions for the joint amplitude both in the spectral and temporal domains, as well as for the chronocyclic Wigner function of heralded single photons, which fully characterizes the single photon spectral (temporal) properties.   On the one hand, we show that unwanted effects of pump chirp in terms of the heralded single photon duration can be suppressed for states designed to be factorable and spectrally elongated.   On the other hand, we show that pump chirp constitutes an effective tool for the control of the degree of photon-pair entanglement.   We show that when frequency-entangled photon pairs propagate through a dispersive medium, entanglement can ``migrate'' between the modulus and phase of the joint temporal amplitude.
\end{abstract}

\pacs{
42.50.Dv, 03.67.Bg}
\maketitle


\section{Introduction}

The ability to generate two-photon states with specific properties is essential for quantum information processing applications.  Specifically, it has been determined that factorability is required for heralding of \textit{pure} single photons\cite{uren05}, a crucial ingredient for quantum computation with linear optics\cite{kok07}.   In previous work, various recipes have been proposed, some of which have been demonstrated experimentally, for the generation of factorable photon pairs through  spontaneous parametric processes \cite{grice01, walton04, raymer05,kuzucu05, uren06,hendrych07,corona07,garay07,mosley08}. In particular, in a recent experiment, we have demonstrated the effectiveness of asymmetric group velocity matching for the preparation of factorable two-photon states, and hence of heralded single photons with a high degree of purity\cite{mosley08,mosley08b}.  In this experiment we have exploited the concept of engineering the photon pairs at the source so as to avoid the need for filtering, thus achieving a high flux of factorable photon pairs.  This scheme and similar ones have one aspect in common:  they require a broadband pump, often one in the form of a train of femtosecond-duration pulses.  Indeed, a monochromatic pump precludes factorability, leading to a state with a high degree of spectral entanglement.   As is also the case with classical optics, the use of femtosecond-duration pulses, greatly increases the vulnerability to dispersive effects.  Naturally, dispersion leads to a significant effect on the resulting photon-pair properties.  It is crucial to understand this effect for the design and implementation of photon pair sources tailored for quantum information processing applications.  In this paper we present a comprehensive study of the effects of dispersion, in particular of second order dispersion, on the properties of photon pairs generated by type-II, collinear parametric downconversion.  We include in our treatment the effects of both: a quadratically chirped pump and propagation of the signal and idler photons through  a medium with second-order dispersion.

Previous work has analyzed some aspects of dispersive effects on photon pairs generated by parametric downconversion.   Thus, for spectrally anti-correlated photon pairs it has been demonstrated that: i)the correlation time remains unaffected if the signal and idler photons propagate through separate dispersive media characterized by opposite-signed quadratic dispersion \cite{franson92}, and ii) the shape of the Hong-Ou-Mandel interference dip is immune to quadratic dispersion experienced by one of the generated photons\cite{steinberg92}.  The latter effect has been proposed as the basis for a quantum optical coherence tomography scheme which benefits from dispersion cancellation\cite{nasr03}.  The temporal broadening of photon pairs due to dispersion which underpins these effects has been been directly observed \cite{valencia02}. Recently, non-local temporal shaping of the signal photon through chirping of the idler photon has been demonstrated\cite{baek08, baek08b}.  The above results rely on a monochromatic pump, and therefore do not consider or exploit chirp in the pump beam.   In Ref.~\cite{erdmann00}, it was shown that two-photon states with positive spectral correlations, produced by a broadband pump, can be immune to dispersion, to all orders.

In the present paper we analyze the interplay of an ultrashort, quadratically chirped pump with quadratic dispersion experienced by the signal and idler photons generated by the process of type-II, collinear PDC.  Our motivation for this work is to analyze  the various dispersive effects which may be relevant in the design of specific sources of photon pairs for quantum information processing applications.  We derive expressions for the resulting joint amplitude, both in the spectral and temporal domains, explicitly including the effects of dispersion.  On the one hand, we show that the negative influence of pump chirp in terms of the attainable herladed single photon temporal duration can be reduced, or even eliminated, through a dispersion suppression effect which occurs for a state designed to be factorable and spectrally-elongated.  On the other hand, we show that pump chirp can in fact be a powerful tool for controlling entanglement in photon pairs.  We study the relationships between: i)the degree of entanglement, ii)the heralded single-photon purity and iii)the expected Hong-Ou-Mandel interference between two photons generated in different crystals.    We study the spectral (temporal) properties of heralded single photons.  Finally, we discuss an effect which occurs during propagation of the PDC photons through a dispersive medium, by which entanglement can ``migrate'' between the modulus and the phase of the joint amplitude.

\section{The two photon state for type-II PDC under the effects of dispersion}
\label{Sec:EfDisp}

In this paper we study the spectral (temporal) properties of photon pairs produced by spontaneous parametric downconversion (PDC) pumped by a train of ultrashort pulses in cases where the pump, signal and idler fields experience dispersion.  In particular, we study the effects on the two-photon state of: i) quadratic chirp in the pump pulses (e.g. due to a dispersive element prior to the crystal),  and ii) propagation of the two photon states after exiting the crystal through a medium with quadratic chirp, such as a fiber.  Throughout this paper, we refer to dispersion introduced by optical elements prior and following the crystal, i.e. excluding dispersion introduced by the crystal itself, as \textit{external} dispersion.  We will concentrate our discussion on collinear, type-II (with the signal and idler photons orthogonally polarized) and frequency degenerate parametric downconversion.

Following a standard perturbative approach, the two-photon state produced can be written as

\begin{equation}
|\Psi\rangle=|\mbox{vac}\rangle+\eta \int d \omega_s \int  d\omega_i f(\omega_i,\omega_s)
|\omega_s\rangle_s |\omega_i\rangle_i,\label{E:2photstate}
\end{equation}

\noindent where we have assumed that the two-photon state is spatially-filtered so that only those $k$ vectors which are parallel to the pump direction of propagation are retained.
Here, $|\omega\rangle_\mu=\hat{a}_\mu
^\dag(\omega)|\mbox{vac}\rangle$ with $\mu=i,s$ where $|\mbox{vac}\rangle$ is the vacuum, $f(\omega_i,\omega_s)$ represents the joint spectral amplitude (JSA) and $\eta$ is a constant related to the conversion effiency.   In the presence of dispersion in addition to spectral filters on the paths of the signal and idler modes, the JSA can be written as

\begin{equation}
f(\omega_i,\omega_s)=\Phi(\omega_i,\omega_s)
\alpha(\omega_i+\omega_s)D(\omega_i,\omega_s)F_i(\omega_i)F_s(\omega_s),
\label{E:JSA}
\end{equation}

\noindent where $\Phi(\omega_i,\omega_s)$ denotes the phasematching function (PMF), $\alpha(\omega_i+\omega_s)$
represents the pump spectral envelope function (PEF), $D(\omega_i,\omega_s)$ describes a phase term
associated with external dispersion, and $F_\mu(\omega)$ (with $\mu=i,s$) describes
a spectral filter acting on each of the signal and idler modes.

The phase matching function can be shown to be given by

\begin{equation}
\Phi(\omega_i,\omega_s)=\mbox{sinc}[L \Delta k(\omega_i,\omega_s)/2]
\mbox{exp}[i L \Delta k(\omega_i,\omega_s)/2],
\label{E:PMF}
\end{equation}

\noindent in terms of the phasemismatch  $\Delta k(\omega_i,\omega_s)=k_p(\omega_s+\omega_i)-k_s(\omega_s)-k_i(\omega_i)$ and the crystal length $L$. In this work we rely both on numerical calculations taking into account dispersion to all orders, as well as on analytical expressions based on a Taylor series description of the phasemismatch.  In the latter case, we can write the phasemismatch as a function of the frequency detunings $\nu_\mu=\omega_\mu-\omega_c$ (with $\mu=s,i$) as

\begin{eqnarray}
L \Delta k(\nu_i,\nu_s) &\approx& L \Delta k^{(0)}+\tau_i
\nu_i+\tau_s \nu_s +b_s \nu_s^2 \nonumber \\
&+& b_i \nu_i^2 + b_p \nu_s \nu_i
\label{E:TE}
\end{eqnarray}

\noindent where perfect phasematching is attained at $\omega_i=\omega_s=\omega_c$, $\Delta k^{(0)}$ is the constant term in the series (which we assume to vanish), and in terms of the following definitions

\begin{eqnarray}
\tau_\mu &=& L[k_p'(2 \omega_c)-k_\mu'(\omega_c)] \\
b_\mu &=& \frac{L}{2}[k_p''(2 \omega_c)-k_\mu''(\omega_c)] \\
b_p &=& L k_p''(2 \omega_c) \label{E:Dispcris}.
\end{eqnarray}

Here, $\tau_\mu$ (with $\mu=s,i$) represents a group velocity mismatch coefficient between the pump and the signal/idler fields, while $b_{p,s,i}$ represent group velocity dispersion coefficients involving the pump, signal and idler frequencies.

The broadband pump is described by the spectral envelope function $\alpha(\omega)$, which we model as a Gaussian function with bandwidth $\sigma$

\begin{equation}
\alpha(\nu_s+\nu_i)=\exp\left[-(\nu_s+\nu_i)^2/\sigma^2\right].
\end{equation}

Likewise, function $F_\mu(\omega)$ (with $\mu=s,i$) is modelled as a Gaussian function with width $\sigma_F$, i.e. $F_\mu(\nu)=\exp(-\nu^2/\sigma_F^2)$.  By neglecting cubic and higher order dispersive terms, and also neglecting linear dispersive terms (which
result in a temporal shift without otherwise modifying the two-photon properties), we can express function
$D(\nu_i,\nu_s)$ as

\begin{equation}
D(\nu_i,\nu_s)=\mbox{exp}\left[i\beta_p(\nu_i+\nu_s)^2\right]\mbox{exp}
\left[i\beta_i\nu_i^2\right]\mbox{exp}\left[i\beta_s \nu_s^2\right]
\label{E:ExDisp}
\end{equation}

\noindent in terms of the group velocity dispersion (GVD) parameters $\beta_\mu$ (with $\mu=p,s,i$).  If these $\beta$ coefficients are due to propagation through a dispersive medium of length $\ell_\mu$ and characterized by wave number $\kappa_\mu(\omega)$ (placed prior to the crystal in the case of the pump and following the crystal in the case of the generated photons), then $\beta_\mu=\ell_\mu \kappa_\mu^{''}/2$, where $''$ denotes a second frequency derivative, evaluated at $\omega_c$ for PDC light and at $2 \omega_c$ for the pump.  Throughout this paper we employ a temporal description of the two-photon state, based on the joint temporal amplitude (JTA) function $\tilde{f}(t_i,t_s)$, obtained as the two-dimensional Fourier transform of the JSA (see Eq.~\ref{E:JSA}).   In terms of the JTA we may write down the two-photon state as:

\begin{equation}
|\Psi\rangle=|\mbox{vac}\rangle+\eta\int dt_s\int dt_i \tilde{f}(t_i,t_s) |t_s\rangle_s |t_i\rangle_i
\end{equation}

\noindent where $|t \rangle_\mu=\tilde{a}^\dagger_\mu(t)|\mbox{vac}\rangle$, defined  in terms of the time-domain annihilation operators

\begin{equation}
\tilde{a}_\mu(t)=\int d \omega \hat{a}_\mu(\omega) e^{-i \omega t}.
\end{equation}

In order to calculate the JTA analytically we use the power series description of the phasemismatch (see Eq.~\ref{E:TE}), up to first order terms.  We also neglect the phase term in the phasematching function (see Eq.~\ref{E:PMF}), which is justified in many of the cases to be analyzed in this paper for which external dispersion dominates over dispersion introduced by the crystal.    The resulting expression is as follows

\begin{align}
\label{E:JTA}
&\tilde{f}({t_i,t_s})=N_t\mbox{exp}\left[-C_2(t_i,t_s)+\frac{C_1(t_i,t_s)^2}{4C_0}\right] \nonumber\\ &\times \bigg(\mbox{erf}\left[\frac{1}{2\sqrt{C_0}}(C_1(t_i,t_s)+2 C_0)\right]  \nonumber\\ & - \mbox{erf}\left[\frac{1}{2\sqrt{C_0}}(C_1(t_i,t_s)-2 C_0)\right] \bigg)
\end{align}

\noindent where $N_t$ is a normalization constant, and which is expressed in terms of the following definitions

\begin{align}
\label{E:JTAc}
C_0 &=-\frac{\tilde{T}_{ss}^2 \tau_i^2 - 2 \tilde{T}_{si}^2 \tau_i \tau_s+ \tilde{T}_{ii}^2 \tau_s^2}{16(\tilde{T}_{si}^4-\tilde{T}_{ii}^2 \tilde{T}_{ss}^2)}\nonumber \\
C_1(t_i,t_s) &=-\frac{t_s (\tilde{T}_{ii}^2 \tau_s - \tilde{T}_{si}^2 \tau_i) + t_i (\tilde{T}_{ss}^2 \tau_i - \tilde{T}_{si}^2 \tau_s)}{4(\tilde{T}_{si}^4-\tilde{T}_{ii}^2 \tilde{T}_{ss}^2)}\nonumber \\
C_2(t_i,t_s) &=-\frac{\tilde{T}_{ss}^2 t_i^2 - 2 \tilde{T}_{si}^2 t_i t_s + \tilde{T}_{ii}^2 t_s^2}{4(\tilde{T}_{si}^4-\tilde{T}_{ii}^2 \tilde{T}_{ss}^2)}.
\end{align}

Here, we have defined parameters $\tilde{T}^2_{\lambda \mu}$ (with $\lambda,\mu =s,i$)  in terms of their real and imaginary parts

\begin{eqnarray}
 \tilde{T}_{\lambda \mu,R}^2 &=&  \delta_{\lambda \mu}/\sigma_F^2+1/\sigma^2 \label{E:T1}\nonumber \\
 \tilde{T}_{\lambda \mu,I}^2 &=&-(\beta_p+\delta_{\lambda \mu} \beta_\mu)
\label{E:T2}
\end{eqnarray}

\noindent where $\delta_{\lambda \mu}$ represents a Kronecker delta. A similar expression, for the case without external dispersion, has been presented in~\cite{mikhailova08}. We also present expressions for the joint amplitude, both in the spectral and temporal domains, based on a Gaussian approximation for the sinc function in the phasematching function.   Clearly, this approximation ignores the sidelobes associated with the sinc function and leads to slightly different spectral properties even near perfect phasematching, compared to the full calculation.   However, the resulting joint amplitude expressions capture the essential physics of the two-photon state and can be exploited for the derivation of  analytic expressions for the following:  i)  Hong-Ou-Mandel interferogram, relying on two independently-generated heralded single photons (Secc.~\ref{Sec:ESPP}), ii) Chronocyclic Wigner function for heralded single photons (Secc.~\ref{Sec:Winger}) , and iii) Schmidt decomposition of the modulus of the joint temporal amplitude (Secc.~\ref{Sec:migration}). Thus, by approximating $\mbox{sinc}(x)\approx \mbox{exp}(-\gamma x^2)$ with $\gamma \approx 0.193$ (this value is selected so that the two functions have identical full widths at half maximum), we can write down the JSA as

\begin{equation}
\label{E:JSAAn}
f(\nu_i,\nu_s)=N_\omega \mbox{exp}\left[-(T_{ii}^2\nu_i^2+T_{ss}^2\nu_s^2+2T_{si}^2\nu_i \nu_s)\right].
\end{equation}

Here, $N_{\omega}$ is a normalization constant, and  parameters $T^2_{\lambda \mu}$ (with $\lambda,\mu =s,i$) are similar to parameters $\tilde{T}^2_{\lambda \mu}$, with an additional term in their real parts

\begin{eqnarray}
T_{\lambda \mu,R}^2 &=&  \tilde{T}_{\lambda \mu,R}^2+\gamma \tau_\lambda \tau_\mu /4 \label{E:T1}\nonumber \\
T_{\lambda \mu,I}^2 &=& \tilde{T}_{\lambda \mu,I}^2
\label{E:T2}.
\end{eqnarray}

The corresponding expression for the JTA is as follows

\begin{equation}
\label{E:JTAapp}
\tilde{f}(t_i,t_s)=N_t \mbox{exp}\left[\Omega_{s}^2 t_i^2+\Omega_{i}^2 t_s^2-2 \Omega_{si}^2 t_i t_s \right],
\end{equation}

\noindent where $N_t$ is a normalization constant and where we have defined the parameters $\Omega_{\mu}$ with $\mu=i,s$ and $\Omega_{si}$ as

\begin{eqnarray}
\Omega_{\mu}^2&=&\frac{T_{\mu \mu}^2}{4(T_{si}^4-T_{ii}^2 T_{ss}^2)} \label{Ommm}\\
\Omega_{si}^2&=&\frac{T_{si}^2}{4(T_{si}^4-T_{ii}^2 T_{ss}^2)}\label{Ommmsi}.
\end{eqnarray}

If the joint amplitude is normalized so that $\int d \nu_s \int d \nu_i |f(\nu_i,\nu_s)|^2=1$, $|f(\nu_i,\nu_s)|^2$ represents a joint probability distribution for the emission of photon pairs with frequency detunings $\nu_s$ and $\nu_i$.  We refer to $|f(\nu_i,\nu_s)|^2$ as the joint spectral intensity, or the joint spectrum.  It is convenient to define, for later use, a correlation coefficient $\Xi=-\sigma_{si}^2/(\sigma_s \sigma_i)$ where $\sigma_{si}^2$ represents the covariance associated with this probability distribution and where $\sigma_\mu$ (with $\mu=s,i$) represents the standard deviation for each of the two marginal distributions.  The correlation coefficient is constrained to take values within the range $-1 \le \Xi \le 1$;  in terms of the parameters defined in Eq.~\ref{E:T1}, the correlation coefficient is given by

\begin{equation}
\Xi=\frac{T_{si,R}^2}{(T^2_{ss,R} T^2_{ii,R})^{1/2}}.
\end{equation}

%
%
%

An important class of two-photon states is composed of those which are unentangled, or \textit{factorable}.  For these states, functions $S(\omega)$ and $I(\omega)$ exist such that the JSA may be expressed as $f(\omega_i,\omega_s)=S(\omega_i)I(\omega_s)$.   Factorable states are required for heralding of \textit{pure} single photons.    In previous work, it has been shown that in cases where a symmetric group velocity matching condition is fulfilled \cite{keller97,grice01}, symmetric and factorable two-photon states, with a round joint spectrum, are possible.  Likewise, it has been shown that in cases where an asymmetric group velocity matching condition is fulfilled, factorable states with an elongated joint spectrum are possible.

For the asymmetric group velocity matching case where $k_p'=k_s'$ or $\tau_s=0$ (i.e. where the pump and signal photons have identical group velocities), the general expression for the JTI (see Eq.~\ref{E:JTA}) may be shown to reduce to

\begin{eqnarray}
|\tilde{f}(t_i,t_s)|^2&=N_{a} \exp \left(-\frac{t_s^2 \sigma^2}{2(1+\beta_p^2 \sigma^4)} \right) \nonumber \\ &\times \mbox{rect} \left(-\frac{\tau_i}{2},\frac{\tau_i}{2};t_s-t_i\right)
\label{E:JTIKDPA}
\end{eqnarray}

\noindent where $N_a$ is a normalization constant.  See the appendix for a definition of the rect function and for more details on how the general form of the joint temporal amplitude (see Eq.~\ref{E:JTA}) simplifies for particular cases.

\begin{figure}[ht]
\begin{center}
\includegraphics[width=3.4in]{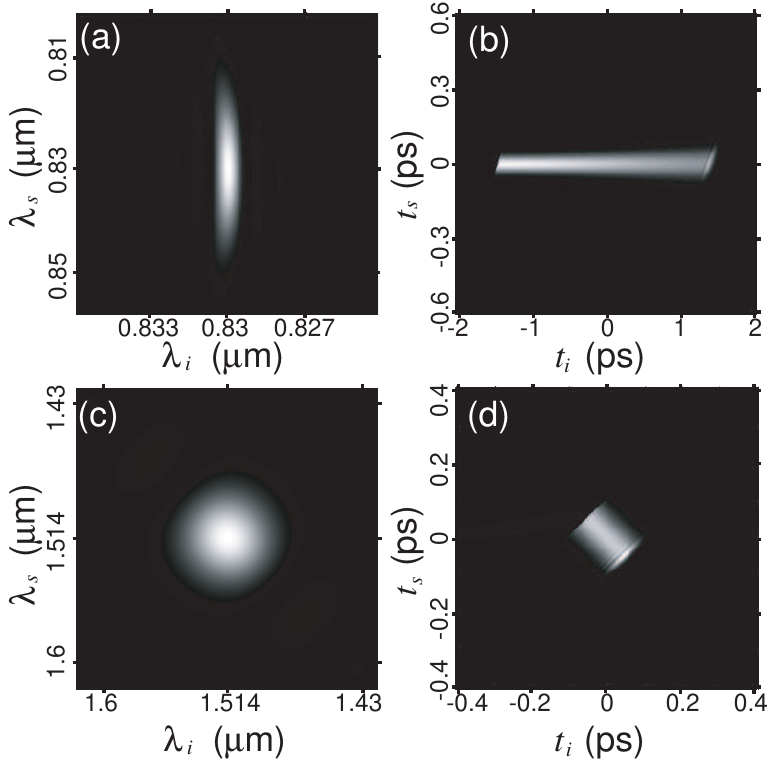}
\end{center}
\caption{(a)  Joint spectral intensity for a nearly-factorable two-photon pair source, which fulfils asymmetric group velocity matching (based on a KDP crystal). (b) Corresponding joint temporal intensity.  (c) Joint spectral intensity for a nearly-factorable two-photon pair source, which fulfils symmetric group velocity matching (based on a BBO crystal). (d) Corresponding joint temporal intensity.}
\label{Fig:JSIJTIKDPBBO}
\end{figure}

Throughout this paper we will refer, as an example of interest, to factorable and spectrally elongated two-photon states obtained by asymmetric group velocity matching (GVM).  Another, related, category of state is that of factorable, symmetric two-photon states obtained by symmetric GVM.   In Fig.~\ref{Fig:JSIJTIKDPBBO} we present, for illustration and future reference, the joint spectral intensity as well as the joint temporal intensity for specific examples of these two categories of source, in the absence of external dispersion; later figures explore the effect of external dispersion.  These plots are derived from numerical calculations, taking into account the full two-photon state with dispersion to all orders (see Eq.~\ref{E:JSA}). In the case of the factorable, asymmetric state we have assumed a pump in the form of an ultrashort pulse train centered at $415$nm with a FWHM bandwidth of $5$nm (which corresponds to $\sigma = 4.65 \times 10^{13} \mbox{rad s}^{-1}$).  Likewise, we have assumed a $2$cm-long potassium dihydrogen phosphate (KDP) crystal with a cut angle of $\theta_{pm}=67.8^\circ$, chosen for type-II, collinear, degenerate phasematching.  In the case of the factorable, symmetric state we have assumed a pump in the form of an ultrashort pulse train centered at $757$nm with a FWHM bandwidth of $15$nm (which corresponds to $\sigma = 4.19 \times 10^{13} \mbox{rad s}^{-1}$).  Likewise, we have assumed a $2.29$mm-long $\beta$ barium borate (BBO) crystal with a cut angle of $\theta_{pm}=28.8^\circ$, chosen for type-II, collinear, degenerate phasematching.

\section{Entanglement and single photon purity}
\label{Sec:ESPP}

\begin{figure}[ht]
\begin{center}
\includegraphics[width=3.1in]{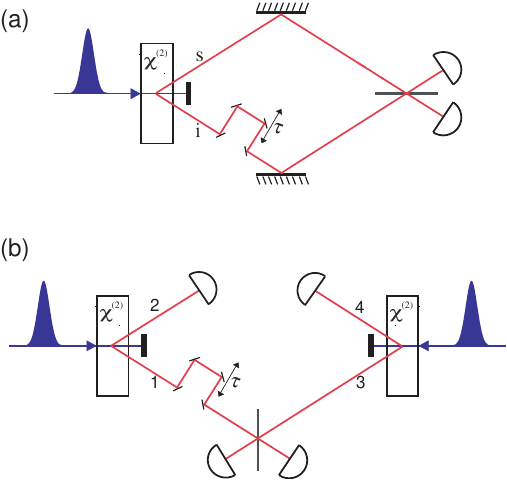}
\end{center}
\caption{(Color online) Schematic diagrams for the single-source [panel (a)] and double-source [panel (b)] Hong-Ou-Mandel interferometer setups.  While these diagrams show, for graphical clarity, PDC sources emitting into non-collinear signal and idler modes, all calculations in our paper refer to type-II collinear sources where the two modes can be distinguished by polarization rather than by optical path.}
\label{Fig:HOMIsetup}
\end{figure}

One way to characterize the spectral (temporal) properties of photon pairs is through their interference properties.  Two photons interfere in a Hong-Ou-Mandel interferometer (HOMI) with perfect visibility if they are indistinguishable.  In the case where the two photons come from the same source [see Fig~\ref{Fig:HOMIsetup}(a)], indistinguishability leads to the requirement that the joint amplitude be symmetric.   In the present analysis, where we concentrate on the spectral degree of freedom, this means specifically that $f(\omega_i,\omega_s)=f(\omega_s,\omega_i)$.  In the case where the two interfering single photons are heralded from independent sources, indistinguishability leads to the requirement that the joint amplitude be factorable.   Concretely, we assume the experimental situation depicted in Fig.~\ref{Fig:HOMIsetup}(b), where two single heralded photons each in the signal mode of two identical sources, are made to interfere at a beam splitter.   We assume that the joint spectral amplitude functions for the two sources are given by the functions $f_1(\omega_i,\omega_s)$ and $f_2(\omega_i,\omega_s)$.  In our analysis, let us begin with the expression for the four-fold coincidence rate as a function of the delay between the two interfering photons in the Hong-Ou-Mandel inteferometer\cite{uren03}

\begin{align}
\label{E:Rcfull}
R_c  \left(\tau \right) &= 1-\int_0^\infty d\omega_1 \int_0^\infty d\omega_2 \int_0^\infty d\omega_3 \int_0^\infty d\omega_4 f_1 \left(\omega_1,\omega_2\right)  \nonumber \\
&\times f_2 \left(\omega_3,\omega_4\right) f_1^* \left(\omega_1,\omega_4\right) f_2^{*} \left(\omega_3,\omega_2\right) \mbox{e}^{i \left(\omega_1-\omega_3\right)\tau}.
\end{align}

Here, we have assumed that each of the joint amplitude functions is normalized.  The subscripts $1$-$4$ in the above expression refer to each of the similary-labelled paths in Fig.~\ref{Fig:HOMIsetup}. Assuming that the minimum coincidence rate (i.e. the lowest point in the interference dip) occurs at zero delay, the interference visibility $V$, defined as the depth of the dip normalized by the background coincidence rate, is given by

\begin{eqnarray}
V&=&1-R_c(0) \nonumber \\
&=& \int_0^\infty d \omega \int_0^\infty d \omega' \rho_{s1}(\omega,\omega') \rho_{s2}(\omega',\omega) \nonumber \\
&=& \mbox{Tr} ( \hat{\rho}_{s1} \hat{\rho}_{s2} ).
\label{E:Vis}
\end{eqnarray}

Here, we have defined the partial density operator $\hat{\rho}_{s\mu}$  for the interfering photon from each of the two sources ($\mu=1,2$), obtained by tracing over idler frequencies; the corresponding matrix elements are

\begin{eqnarray}
\rho_{s \mu}(\omega,\omega')&=&\langle \omega | \mbox{Tr}_{i}(\hat{\rho}_\mu) | \omega' \rangle \nonumber \\
&=&  \int_0^\infty d \omega_0 f_\mu(\omega_0,\omega) f^*_\mu(\omega_0,\omega')
\end{eqnarray}

\noindent in terms of the two-photon density operators $\hat{\rho}_\mu=| \Psi_\mu \rangle \langle \Psi_\mu |$ and the partial trace over idler frequencies $\mbox{Tr}_i$.  If we assume that the two sources are in fact identical then, from Eq.~\ref{E:Vis}, $V=\mbox{Tr}(\hat{\rho}_s^2)$ which represents the purity of the heralded, interfering single photons.  Therefore, from an experimental measurement of the two-source Hong-Ou-Mandel interference dip, we may infer the purity of the interfering single photons.

Let us now express the joint amplitudes (assumed to be the same for both sources) in terms of a Schmidt decomposition \cite{law00}

\begin{equation}
\label{E:Schmidt}
f(\omega_i,\omega_s)=\sum_{n} \sqrt{\lambda_n} u_n(\omega_i)v_n(\omega_s)
\end{equation}

\noindent where $\lambda_n$ are the Schmidt eigenvalues  and $u_n(\omega_i)$ and $v_n(\omega_s)$ are the Schmidt functions.  Note, that the Schmidt decomposition can also be performed in the temporal domain, in terms of the corresponding eigenvalues $\lambda_n^{(t)}$ and temporal Schmidt functions $u_n^{(t)}(t_i)$ and $v_n^{(t)}(t_s)$

\begin{equation}
f(t_i,t_s)=\sum_{n} \sqrt{\lambda_n^{(t)}} u_n^{(t)}(t_i)v_n^{(t)}(t_s).
\end{equation}

It is straightforward to show that the reduced density matrix elements can be expressed in terms of the spectral Schmidt functions as

\begin{equation}
\rho(\omega,\omega')=\sum_n \lambda_n v_n(\omega) v_n^*(\omega'),
\label{E:Densopsum}
\end{equation}

\noindent which in general represents a mixed state.  Calculating the visibility from Eq.~\ref{E:Vis}, assuming that both interfering single heralded photons are identical ($\hat{\rho}_{s1}=\hat{\rho}_{s2}\equiv\hat{\rho}$), we can furthermore show that

\begin{eqnarray}
V=\mbox{Tr}(\hat{\rho}^2)=\sum_n \lambda_n^2 \equiv K^{-1}.
\label{E:VK}
\end{eqnarray}

Here we have exploited the orthogonality of the Schmidt functions $u_n(\omega)$.  As may be appreciated from Eq.~\ref{E:VK}, the HOMI visibility in the specific case where the two interfering photons are identical corresponds to reciprocal of the Schmidt number $K$. (The Schmidt number $K$ can be used to quantify the degree of entanglement; while $K=1$ represents a factorable, i.e. unentangled state, $K \gg 1$ represents a highly entangled state).  Therefore, according to Eq.~\ref{E:VK}, from an experimental measurement of the two-source Hong-Ou-Mandel interference dip, we may likewise infer the degree of entanglement of the two photon states.  Let us note that in the derivation of Eq.~\ref{E:VK} we have assumed an ideal situation for which possible experimental imperfections can be disregarded.   In a realistic situation, Hong-Ou-Mandel interference will be influenced by many experimental factors, with the implication that it will no longer be possible to infer the degree of entanglement from the HOMI visibility.  In this case, however, a lower bound for the single-photon purity or an upper bound for the degree of entanglement (rather than the actual values) can be inferred from the Hong-Ou-Mandel measurement.

From the above discussion it is clear that the generation of pure, heralded single photons requires a source factorable photon pairs, i.e. one for which there is a single term in the sum of Eq.~\ref{E:Schmidt} (and therefore, also in the sum of Eq.~\ref{E:Densopsum}).

For the two photon state expressed in terms of the phasemismatch up to first order and in terms of the Gaussian approximation (see Eq.~\ref{E:JSAAn}), the four-fold coincidence rate for the two-source HOMI arrangement is given by

\begin{equation}
R_c(\tau ) = 1- V \mbox{exp}\big(-\tau^2/\Delta \tau^2 \big)
\label{E:Crate}
\end{equation}

\noindent in terms of the visibility $V$

\begin{equation}
V  =\left( \frac{ T_{ii,R}^2 T_{ss,R}^2-(T_{si,R}^2)^2}{(T_{si,I}^2)^2 +T_{ii,R}^2T_{ss,R}^2}\right)^{\frac{1}{2}}
\label{E:Visexp}
\end{equation}

\noindent and the dip temporal width $\Delta \tau$

\begin{align}
\Delta \tau^2  =4\frac{(T_{si,I}^2)^2 +T_{ii,R}^2T_{ss,R}^2}{T_{ii,R}^2}.
\label{E:twidip}
\end{align}

Note that according to Eq.~\ref{E:VK}, there is a reciprocal relationship between $V$ and $K$ so that the Schmidt number may be determined as $K=1/V$ from Eq.~\ref{E:Visexp}.  It may be seen that the visibility reaches unity, as expected, if the mixed term responsible for spectral entanglement $T^2_{si}$ vanishes.  Note that while the dip shape (visibility and width) exhibits a dependence on the pump chirp parameter, it is  independent of the dispersion experienced by the signal and idler photons after exiting the crystal.  This is as expected, since the dip shape is a signature of the type and degree of entanglement present, which cannot change through lossless propagation of the signal and idler photons after exiting the crystal.

\begin{figure}[ht]
\begin{center}
\includegraphics[width=2.7in]{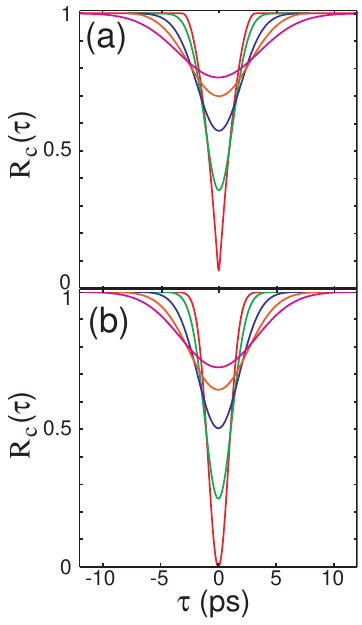}
\end{center}
\caption{(Color online) Fourth-fold coincidences (normalized to a unit background) vs temporal delay between two separate heralded single photons interfering through a Hong-Ou-Mandel interferometer.  These curves result from: (a) a numerical simulation, based on Eq.~\ref{E:Rcfull}, and (b)  the analytic expression in Eq.~\ref{E:Crate}.  The different curves in each panel correspond to the following levels of pump chirp: i)$\beta_p=0$ (shown in red), ii) $\beta_p=1.19\times 10^{-26} s^2$ (shown in green), iii) $\beta_p=2.39\times10^{-26} s^2$ (shown in blue), iv) $\beta_p=3.58\times10^{-26} s^2$ (shown in orange) and v) $\beta_p=4.77\times10^{-26} s^2$ (shown in magenta); in this family of curves, lower visibilities correspond to higher levels of pump chirp.}
\label{Fig:HOMI}
\end{figure}

We illustrate this behavior in Fig.~\ref{Fig:HOMI}, in the context of a factorable, spectrally elongated source based on a KDP crystal (with the same parameters as assumed for Fig.~\ref{Fig:JSIJTIKDPBBO}(a)).   In particular, we show the expected HOMI dip for four different levels of pump chirp (see figure caption).  While the dips shown in Fig.~\ref{Fig:HOMI}(a) were calculated numerically from the full two-photon state, those in Fig.~\ref{Fig:HOMI}(b) were calculated through the analytic expression in Eqns.~\ref{E:Crate} through \ref{E:twidip}.  In general terms, the greater structure in the joint spectrum, e.g. derived from the sinc function, in the full calculation results in lower visibilities compared to the approximate analytic calculation.
As expected, a larger degree of pump chirp leads to lower HOMI visibilities, signalling a greater degree of entanglement in the two-photon states. In particular, no pump chirp leads to perfect visibility (according to the approximate calculation) and to the highest attainable visibility ($V=94 \%$) in the case of the full calculation.

In the next section, we further explore the important relationship between pump chirp and the resulting degree of entanglement.

\section{Control of entanglement: use of a chirped pump}

The information pertaining to dispersion in the two-photon state, as may be seen from Eqns.~\ref{E:JSA} and \ref{E:ExDisp}, is contained in phase terms.   Note that the phase introduced by propagation of the signal and idler photons through dispersive media is factorable into separate signal and idler contributions and therefore does not contribute to entanglement. In contrast, the phase term associated with a chirped pump (e.g. due to transmission of the pump beam through a dispersive medium before reaching the crystal), in general cannot be factored into signal and idler factors and therefore does contribute to entanglement.  This contribution of pump chirp to entanglement, in the spectral domain is necessarily in the form of phase entanglement\cite{chan04}, and has no effect on the joint spectral intensity.  However, phase entanglement in the spectral domain can, as we will discuss, translate into explicit correlations in the joint \textit{temporal} intensity.  Of course, apart from external dispersion considered in this paper, other experimental parameters such as crystal length and pump bandwidth can have a profound effect on the resulting degree of entanglement; these effects have been studied elsewhere (see for example Refs.~\cite{grice01} and \cite{mikhailova08}).

\begin{figure}[ht!]
\begin{center}
\includegraphics[width=2.8in]{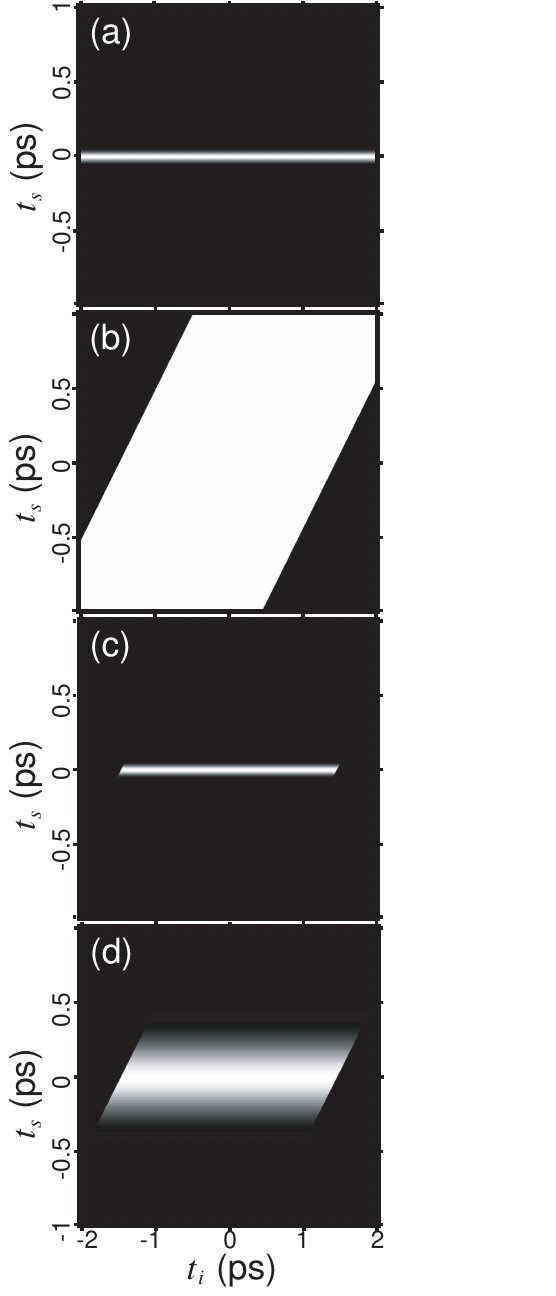}
\end{center}
\caption{This figure shows the effect of pump chirp on the temporal properties of the two-photon state.  In particular we show, plotted as a function of the signal and idler times of emission: (a) the Gaussian function in Eq.~\ref{E:JTA}, for the case of no pump chirp, (b) the rect function in Eq.~\ref{E:JTA}, (c) the square modulus of the product of the functions in the previous two panels, (d) same as in (c), but for a chirped pump with $\beta_p=3.04  \times 10^{-27}s^2$.}
\label{Fig:JTIKDPAna}
\end{figure}

The phase term due to pump chirp, linear in the sum of frequencies $\omega_i+\omega_s$, in general manifests itself as an elongation of the joint temporal intensity, along the direction given by $t_i+t_s$, when plotted in the times of emission space $\{t_i,t_s\}$.  We illustrate the effect of pump chirp, presenting as an example the case of nearly-factorable two-photon states (in the absence of pump chirp) obtained through asymmetric group velocity matching.  In Fig.~\ref{Fig:JTIKDPAna}, we present the joint temporal intensity expected for a KDP crystal, with the same parameters as assumed for Fig.~\ref{Fig:JSIJTIKDPBBO}(a).  In this case, the pump group velocity equals that of the signal (ordinary-wave) photon.  Fig.~\ref{Fig:JTIKDPAna}(a) shows a plot of the Gaussian term in Eq.~\ref{E:JTIKDPA} vs the times of emission $t_i$ and $t_s$, assuming $\beta_p=0$.  Fig.~\ref{Fig:JTIKDPAna}(b) shows a corresponding plot of the rect function in Eq.~\ref{E:JTIKDPA}.  Fig.~\ref{Fig:JTIKDPAna}(c) shows the joint temporal intensity $|\tilde{f}(t_i,t_s)|^2$ in the case of no pump chirp, obtained from the product of the functions plotted in the previous two panels.  From Eq.~\ref{E:JTIKDPA}, the effect of pump chirp becomes, clear:  the width of the Gaussian function in Fig.~\ref{Fig:JTIKDPAna}(b) increases, thus revealing more of the diagonal structure provided by the rect function in the joint temporal intensity.   This behavior is clear from Fig.~\ref{Fig:JTIKDPAna}(d), where we have plotted the JTI  for a pump chirp parameter $\beta_p=3.04 \times 10^{-27}$s$^2$ (corresponding to propagation through a $6$cm thickness of fused silica).

The phase term in the phasematching function (see Eq.~\ref{E:PMF}), which is not taken into account in the analytic expressions of section \ref{Sec:EfDisp} (i.e. Eqns.~\ref{E:JTA} and \ref{E:JSAAn}), in general can contribute to the entanglement in the two-photon state.  Indeed, the mixed term proportional to $\nu_s \nu_i$ in Eq.~\ref{E:TE} leads to a phase term in Eq.~\ref{E:JSAAn} which cannot be factored into signal and idler contributions.  This small effect can be graphically appreciated by comparing Fig.~\ref{Fig:JSIJTIKDPBBO}(b), which takes into account numerically the crystal phase term in the JSA, with Fig.~\ref{Fig:JTIKDPAna}(c) calculated analytically from Eq.~\ref{E:JTIKDPA}, which does not take into account the crystal phase term in the JSA.  We can see that portions of the JTI involving higher values $t_i$, emitted near the second crystal face, involve a broader signal photon, relative to portions of the JTI involving lower values of $t_i$.  This is a consequence of temporal elongation of the pump pulse (which leads to a corresponding signal mode temporal elongation) as it propagates through the crystal.  This intra-crystal dispersion and its effect on the resulting degree of entanglement, may be compensated for by pump chirp, involving dispersion opposite in sign to that experienced by the pump beam in the non-linear crystal; this technique involving making $b_p=-4 \beta_p$ (parameter $b_p$ defined in Eq.~\ref{E:Dispcris}) was discussed in Ref.~\cite{uren05}.   In practice, however, the effect of the dispersive phase associated with the crystal, on the degree of entanglement tends to be small.  For example, for the KDP source discussed above, the value of the Schmidt number, drops from $K=1.065$ to $K=1.061$, under the effects of chirp compensation.

\begin{figure*}[ht]
\begin{center}
\includegraphics[width=6.8in]{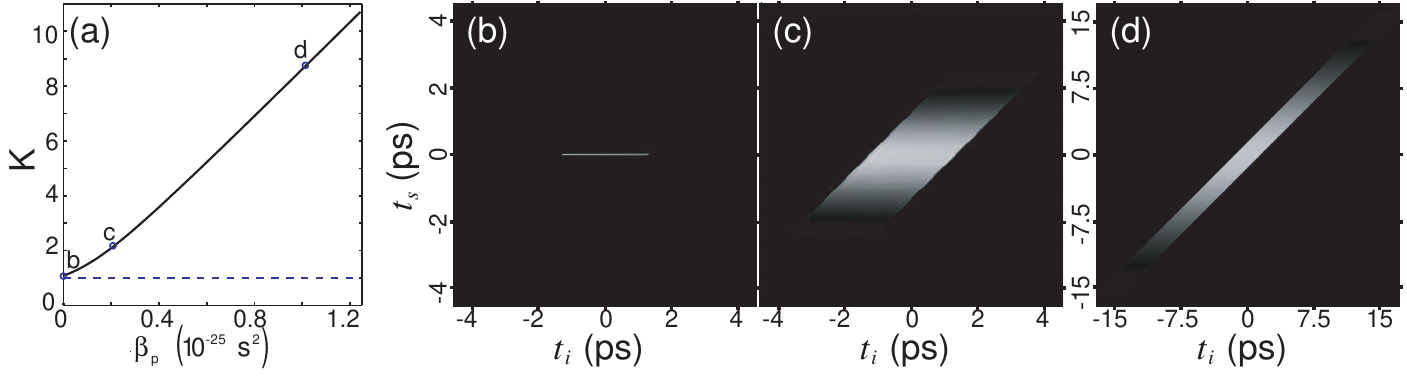}
\end{center}
\caption{(a)This plot shows the behavior of the Schmidt number, which quantifies the degree of entanglement, as a function of the pump chirp parameter, for the same source as in Fig.~\ref{Fig:JSIJTIKDPBBO}(a) and (b); this illustrates that pump chirp may be used as a powerful tool to control the degree of entanglement. Panels (b)-(d) show the joint temporal intensity, for three different values of pump chirp: $\beta_p=0 s^2$, $\beta_p=2.07\times10^{-26} s^2$ and $\beta_p=1.01\times10^{-25} s^2$, identified along the curve in panel (a).}
\label{Fig:ChirpKDP}
\end{figure*}

In addition to the chirp compensation described above, pump chirp can be used as an effective tool to control the degree of entanglement in PDC two-photon sources.  The effect of  pump chirp on the JTI may be appreciated in Figs.~\ref{Fig:ChirpKDP}(c) and (d). As the pump chirp parameter is increased, the correlations apparent in the JTI become more pronounced.   If the two-photon state is factorable in the absence of dispersion, then by adding a controlled level of chirp to the pump pulses (prior to the crystal), it becomes possible to generate two-photon states with an arbitrary degree of entanglement, determined by the pump chirp $\beta_p$.  In order to illustrate this behavior, Fig.~\ref{Fig:ChirpKDP}(a) shows, for a source similar to that assumed for Fig.~\ref{Fig:JSIJTIKDPBBO}(b) the Schmidt number, calculated numerically, as a function of the pump chirp parameter $\beta_p$ (where we have assumed no dispersion experienced by the signal/idler photons).  We can see that the Schmidt number exhibits a monotonically increasing dependence on the pump chirp parameter.  Figs.~\ref{Fig:ChirpKDP}(b) through (d) show the JTI calculated numerically for three different values of the pump chirp, identified along the $K$ vs $\beta_p$ plot in Fig.~\ref{Fig:ChirpKDP}(a).  Of course, according to Eqns.~\ref{E:VK}, a greater degree of entanglement as quantified by $K$, translates into a lower single heralded photon purity, and a lower visibility in a two-source HOMI experiment, as demanded by the conclusions of Sec.~\ref{Sec:ESPP}.

The entanglement contributed to the photon pair by pump chirp in general manifests itself in the form of modified temporal correlations.  In the specific case of Fig.~\ref{Fig:ChirpKDP}, the strength of temporal correlations increases monotonically with the pump chirp parameter $\beta_p$.  These correlations may be quantified for example through the so-called Fedorov ratio\cite{fedorov04}, here defined in the temporal domain as

\begin{equation}
\mathcal{F}=\frac{\Delta t_S}{\Delta t_C (t_0)}\label{E:Fedorov}
\end{equation}

\noindent given in terms of $\Delta t_C (t_0)$ which represents the signal-mode temporal width conditioned on a certain idler detection time $t_0$  and the unconditional signal-mode temporal width $\Delta t_S$.  Thus, while $\Delta t_C (t_0)$ is defined as the full width at half maximum of the function $f_i(t_i)=|f(t_i,t_0)|^2$ for fixed $t_0$ selected to maximize the probability of emission, $\Delta t_S$ is defined as the full width at half maximum of the marginal distribution $\int d t_s |f(t_i,t_s)|^2$.  A numerical calculation of the Fedorov ratio for the situations depicted in Fig.~\ref{Fig:ChirpKDP}(b)-(d) yields the values $1.06$, $1.14$ and $3.98$ respectively.

\section{Spectral (temporal) properties of heralded single photons}
\label{Sec:Winger}

A key application of spontaneous PDC is the generation of heralded single photons.  Here we consider how the continuous variable entanglement properties of the photon pair translate into the properties of the heralded single photons, in the presence of dispersive elements on the paths of the pump, signal and idler.  The spectral (temporal) properties of signal photons heralded by an idler detection event may be conveniently characterized in terms of the heralded single photon chronocyclic Wigner function (CWF). As shown in Ref.~\cite{uren07}, the CWF for PDC light characterized by a JSA $f(\nu_i,\nu_s)$ is given by

\begin{align}
W_s(\nu,t)&=\int d\nu_0 g(\nu_0) \int d \omega'
f\left(\nu_0,\nu+\frac{\omega'}{2} \right)  \nonumber \\
&\times f^*\left(\nu_0,\nu-\frac{\omega'}{2} \right) e^{i \omega' t},
\label{E:CWF}
\end{align}

\noindent where $g(\nu)$ represents the idler detection efficiency. The CWF is a real-valued function which represents a quasi-probability distribution for the emission of a single photon with frequency detuning $\nu$ and at time $t$; the marginal distributions of the CWF yield the corresponding spectral and temporal single-photon intensity profiles.  Here
we generalize the result in Ref.~\cite{uren07} to the case where the three fields involved experience dispersion. We assume that the idler detection efficiency has a Gaussian spectral shape $g(\nu)=\exp[-\nu^2/\sigma_g^2]$ (centered at $\omega=\omega_{c}$ and with bandwidth $\sigma_g$). Carrying out the integrals in Eq.~\ref{E:CWF} with the JSA given in terms of a linear approximation for the phase mismatch (see Eq.~\ref{E:TE}) and using the Gaussian approximation for the phasematching function (see text before Eq.~\ref{E:JSAAn}), we obtain

\begin{align}
W_s(\nu,t) &=\frac{\sqrt{4-\Gamma^2 \Delta t^2 \Delta \omega^2}}{2\pi \Delta t \Delta \omega}
\mbox{exp}\left[-\frac{\nu^2}{\Delta
\omega^2}\right] \mbox{exp}\left[-\frac{t^2}{\Delta t^2}\right] \nonumber \\
& \times \mbox{exp}\left[\Gamma \nu t\right],
\label{E:CWFAna}
\end{align}

\noindent in terms of a mixed-term coefficient $\Gamma$, the temporal duration $\Delta t$ and the spectral width $\Delta
\omega$

\begin{eqnarray}
\Gamma &=&   \frac{2 T_2^2}{T_1^2} \nonumber \\
\Delta t^2 &=& T_1^2 \\
\Delta \omega^2 &=& \frac{T_1^2}{T_2^4+T_1^2 T_3^2}\nonumber ,
\end{eqnarray}

\noindent which are expressed, in turn, in terms of the following definitions

\begin{eqnarray}
T_1^2&=&\frac{2}{\hat{T}_{ii,R}^2}({T_{ss,R}^2} {\hat{T}_{ii,R}^2}+({T_{si,I}^2})^2) \nonumber \\
T_2^2&=&\frac{2}{\hat{T}_{ii,R}^2}({T_{ss,I}^2}{\hat{T}_{ii,R}^2}-{T_{si,R}^2}{T_{si,I}^2})\nonumber\\
T_3^2&=&\frac{2}{\hat{T}_{ii,R}^2}({T_{ss,R}^2} {\hat{T}_{ii,R}^2}-({T_{si,R}^2})^2).
\label{E:WgTs}
\end{eqnarray}

Here, we have redefined $T_{ii}$, now referred to as $\hat{T}_{ii}$, to include the effect of the idler detection efficiency; it is expressed as follows

\begin{equation}
\hat{T}^2_{ii}=T^2_{ii}+\frac{1}{\sigma_g^2}.
\end{equation}

Interestingly, dispersion experienced by the idler photon does not have an effect on the spectral (temporal) properties of a heralded single photon in the signal mode; indeed Eqns.~\ref{E:WgTs} exhibit no dependence on $\beta_i$.   Note also that in the absence of external dispersion, i.e. if $\beta_s=\beta_i=\beta_p=0$, $T_1^2$ reduces to $T_1^2=2 {T^2_{ss}}_{,R}$, and $T_2^2=0$; on the other hand, $T_3^2$ remains unaffected by the presence or absence of dispersion.  This in turn implies that, according to Eqns.~\ref{E:WgTs}, in the absence of dispersion,  the mixed term $\Gamma$ vanishes,  in which case we recover the expression for the CWF in Ref.~\cite{uren07}.  The effect of this mixed term, due to dispersion, is that the CWF becomes tilted in $\{\omega,t\}$ space in such a way that the frequency marginal distribution remains unaffected and the temporal marginal distribution is broadened.  In order to see this explicitly, let us obtain expressions for the marginal distributions.  The spectral intensity profile of the heralded single photon is given by

\begin{eqnarray}
\label{ }
&&I_\nu(\nu)=\int dt W_s(\nu,t)\nonumber  \\ &&=\left(\frac{4-\Gamma^2 \Delta t^2 \Delta \omega^2}{4 \pi \Delta \omega^2}\right)^{1/2}
    \mbox{exp}\left[-\frac{\nu^2}{\Delta \omega_M^2}\right],
\end{eqnarray}

\noindent in terms of the single-photon spectral width $\Delta \omega_M$

\begin{eqnarray}
\Delta \omega_M^2=\frac{1}{T_3^2}.
\label{wspect}
\end{eqnarray}

The temporal intensity profile of the heralded single photon is given by

\begin{eqnarray}
\label{ }
&&I_t(t)=\int d \nu W_s(\nu,t)\nonumber\\ &&=\left(\frac{4-\Gamma^2 \Delta t^2 \Delta \omega^2}{4 \pi \Delta t^2}\right)^{1/2}
 \mbox{exp}\left[-\frac{t^2}{\Delta t_M^2}\right],
\end{eqnarray}

\noindent in terms of the single-photon temporal width $\Delta t_M$

\begin{equation}
 \Delta t_M^2=T_1^2+\frac{T_2^4 }{T_3^2}.
 \label{wtemp}
\end{equation}

As expected, while the single-photon spectral width $\Delta \omega_M$ exhibits no dependence on the external dispersion, in general the single-photon temporal width $\Delta t_M$ is broadened due to dispersion.  In order to further understand this broadening, it is convenient to express the single photon temporal duration as $\Delta t_M^2=\Delta t_0^2+\Delta$ where $\Delta t_0^2=2 {T^2_{ss}}_{,R}$ represents the single photon duration for no dispersion.   We can express $\Delta$ as

\begin{equation}
\frac{\Delta}{2}=\frac{{T^2_{ii}}_{,R} ({T^2_{ss}}_{,I})^2+{T^2_{ss}}_{,R}({T^2_{si}}_{,I})^2- 2 {T^2_{si}}_{,R} {T^2_{ss}}_{,I} {T^2_{si}}_{,I}}{{T^2_{ss}}_{,R} {T^2_{ii}}_{,R} -({T^2_{si}}_{,R})^2}.
\label{E:delt}
\end{equation}

Re-writing the above expression in terms of the correlation coefficient $\Xi$, and using: i)the fact that $\Xi$ is constrained by $-1 \le \Xi \le 1$ and ii) the inequality $x^2+y^2-2 x y \xi \ge 0$, valid for all $x,y$ with $-1 \le \xi \le 1$,  we can easily show that $\Delta \ge 0$, so that $\Delta t_0$ represents the shortest possible single-photon temporal duration. In other words, as expected, the shortest single-photon temporal duration occurs for no dispersion in any of the three optical fields.

\begin{figure}[ht]
\begin{center}
\includegraphics[width=3.2in]{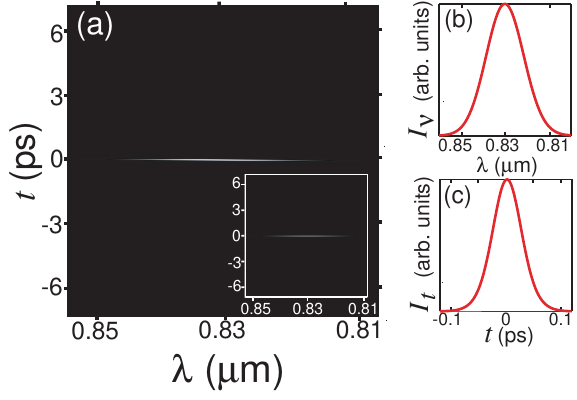}
\end{center}
\caption{(a) Chronocyclic Wigner function plotted for a source identical to that assumed for Fig.~\ref{Fig:JSIJTIKDPBBO}(a) and (b), computed numerically from Eq.~\ref{E:CWF}. Inset: plot resulting from the analytic expression in Eq.~\ref{E:CWFAna}. (b)  Spectral intensity of heralded single photon, in the signal mode, computed as one of the marginal distributions of the CWF.  (c)  Temporal intensity of heralded single photon, in the signal mode, computed as one of the marginal distributions of the CWF. }
\label{Fig:Wig0D}
\end{figure}

In order to illustrate the characterization of spectral (temporal) heralded single photon properties, Fig.~\ref{Fig:Wig0D} shows a plot of the resulting chronocyclic Wigner function for the source parameters assumed in Fig.~\ref{Fig:JSIJTIKDPBBO}~(a), and assuming no dispersion experienced by the pump, signal and idler modes, no PDC filtering ($\sigma_F \rightarrow \infty$) and ideal triggering ($\sigma_g \rightarrow \infty$).
While the main plot in panel (a) shows the numerically-obtained CWF (through numerical integration of Eq.~\ref{E:CWF}), the inset shows the CWF plotted from our analytic expression (Eq.~\ref{E:CWFAna}).    Panels (b) and (c) show the CWF marginal distributions, equivalent to the spectral and temporal single-photon intensity profiles.   The effects of dispersion on the heralded single-photon properties will be discussed in the following section.

\section{Dispersion suppression effects}

The factorable, spectrally elongated state leads to some interesting dispersion suppression properties.  As was already discussed in Ref.~\cite{uren05}, in the limit of an ``ideal'' factorable, spectrally elongated state where the signal photon is monochromatic, i.e. where the JSA can be expressed as $f(\nu_i,\nu_s)=\delta(\nu_s)I(\nu_i)$, the two-photon state remains factorable despite the presence of dispersion.  Indeed, in this case the JSA can then be expressed as:

\begin{eqnarray}
f(\nu_i,\nu_s)&=&\delta(\nu_s)I(\nu_i) \exp[i(\beta_p+\beta_s) \nu_s^2]\nonumber \\ &\times&
 \exp[i(\beta_p+\beta_i) \nu_i^2] \exp[i 2 \beta_p \nu_s\nu_i].
\label{E:JSAdelta}
\end{eqnarray}

The above expression includes a phase term which is not factorable into signal and idler contributions.  However, the effect of this phase term which depends on the product $\nu_s \nu_i$, and which is controlled by the pump chirp parameter $\beta_p$,  is suppressed in the limiting case where one of the two photons produced is monochromatic; dispersive effects fundamentally require a bandwidth.  Note, however, that while this implies that an ideal factorable two-photon state remains factorable despite an arbitrary level of pump chirp, the two photons do individually experience temporal broadening due to the phase terms proportional to $\nu_s^2$ and $\nu_i^2$.  We can also see from Eq.~\ref{E:JSAdelta} that if pump chirp is compensated by an equal magnitude of signal and idler chirp, but with the opposite sign, then the two photon state becomes unaffected by the presence of external dispersion, since in this case phase terms are either not present or yield no effect on the two-photon state.

In order to study these effects for a realistic state, i.e. one where the idler photon has a small, but non-zero bandwidth,
let us use our expressions for the two-photon state based on the linear dispersion, and Gaussian approximations.  We will begin this analysis by writing the joint spectral amplitude in terms of the adimensional detunings defined by $n_\mu= (T^2_{\mu \mu,R})^\frac{1}{2} \nu_\mu$ (with $\mu=s,i$)

\begin{eqnarray}
&f_n(n_i,n_s)=N_\omega \exp \big[- n_s^2- n_i^2 - 2 \Xi n_s n_i  \nonumber \\
&-i \frac{T^2_{ss,I}}{T^2_{ss,R}} n_s^2 - i \frac{T^2_{ii,I}}{T^2_{ii,R}} n_i^2
-2 i \frac{T^2_{si,I}}{(T^2_{ss,R} T^2_{ii,R})^\frac{1}{2}} n_s n_i \big].
\end{eqnarray}

For the two-photon state to be factorable, the two mixed terms (one real and one imaginary) proportional to $n_s n_i$ must be vanishingly small. In the absence of external dispersion, a factorable, asymmetric two-photon state can be obtained if: i) an asymmetric group velocity matching condition is fulfilled (leading to $\tau_s=0$), and ii) for a relatively long crystal, coupled with a relatively large pump bandwidth, such that $|\sigma \tau_i| \gg 1$.  Under these conditions, it is straightforward to show that the real parts of the $T$ coefficients which define the JSA (see Eq.~\ref{E:T2}) may be expressed as $T^2_{ii,R}=\gamma \tau_i^2/4$, $T^2_{ss,R}=1/\sigma^2$ and $T^2_{si,R}=1/\sigma^2$.  Thus,  the correlation coefficient becomes $\Xi=2/(\sqrt{\gamma} \sigma \tau_i)$ which vanishes for the conditions which define this state.  The coefficient which defines the imaginary mixed term may be written as $(2 \sigma^2 \beta_p)/(\sqrt{\gamma} \sigma \tau_i)$.  Thus, a state fulfilling asymmetric group velocity matching becomes factorable in the presence of pump chirp if in addition to $|\sigma \tau_i| \gg 1$, the condition $|\sigma \tau_i| \gg| \sigma^2 \beta_p|$ is also fulfilled.  This last condition tells us that if the adimensional group velocity mismatch  $\sigma \tau_i$ greatly exceeds the adimensional pump chirp $\sigma^2 \beta_p$, the two photon state remains factorable despite the presence of pump chirp.  Clearly, the larger the coefficient $\tau_i$ (linearly proportional to the crystal length), the more pump chirp can be present while retaining two-photon factorability. Also, note that this ``immunity'' to pump chirp is weakened for larger values of the pump bandwidth $\sigma$.  This effect makes factorable and spectrally elongated states, for which the above dispersion suppression effect occurs, attractive for practical implementations of quantum information processing protocols, since in most realistic experimental situations there is some chirp present in the pump beam which under typical conditions would suppress factorability.

Recalling that $T^2_{ss,I}=-(\beta_p+\beta_s)$ (and similarly for the corresponding idler quantity), we can see from Eq.~\ref{E:JSAdelta} that if $-\beta_p=\beta_s=\beta_i$, in addition to the fulfilment of the conditions from the previous paragraph, then the two-photon state becomes completely immune to external dispersion.  In particular, the temporal duration of the signal photon in the presence of chirp becomes identical to the corresponding duration without external dispersion.   In order to see this,  it is clear from Eq.~\ref{E:T2} that if $\beta_p=-\beta_s$ leading to $T^2_{ss,I}=0$, then $\Delta/\beta_p =(\sigma^2 \beta_p)/(\gamma \sigma^2 \tau_i^2/4-1)$ (see Eq.~\ref{E:delt}).  Thus, if $|\sigma \tau_i| \gg 1$ and $|\sigma^2 \beta_p| \ll |\sigma \tau_i|$ (the conditions discussed in the previous paragraph), $\Delta \rightarrow 0$ and therefore the signal temporal duration becomes equal to that in the absence of external dispersion.

\begin{figure*}[ht!]
\begin{center}
\includegraphics[width=6.0in]{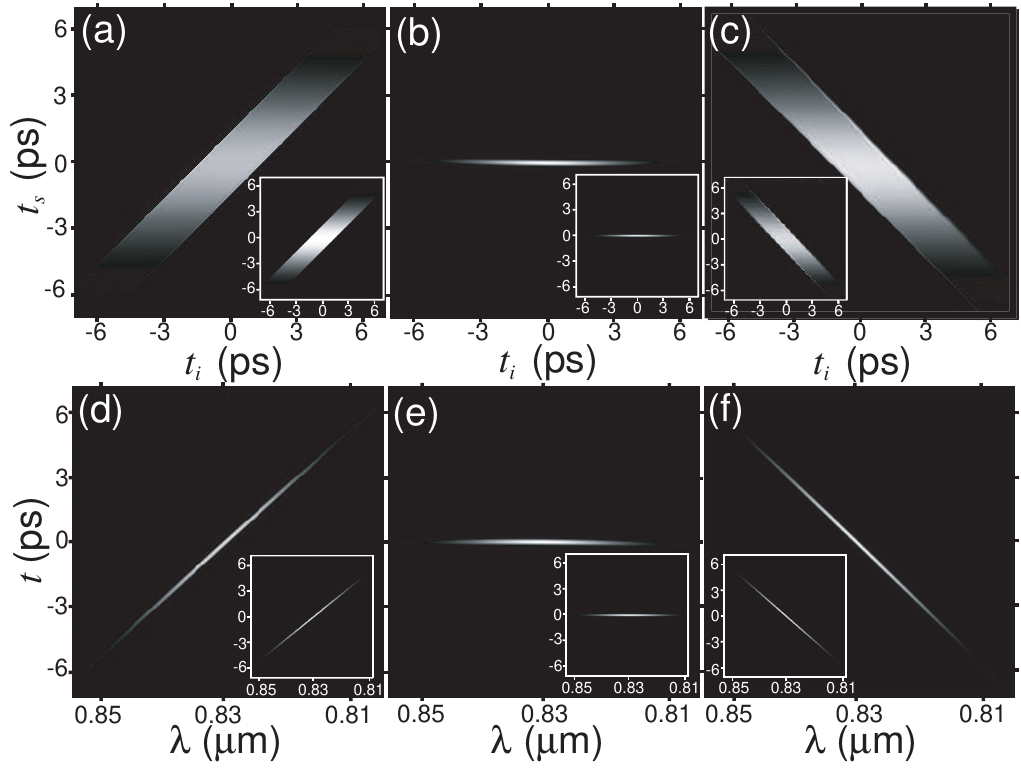}
\end{center}
\caption{Panels (a)-(c):  Joint temporal intensity for three different levels of chirp [$\beta=0$,$\beta=-\beta_p$ and $\beta=-2 \beta_p$, corresponding to the following lengths of propagation through fused silica fiber: $L=0$m, $L=2.64$m and $L=5.28$m] experienced by the signal and idler photons, and a pump chirp level of $\beta_p=-4.77 \times 10^{-26}s^2$.  While these plots were computed numerically through Eq.~\ref{E:JSA}, the insets show the corresponding plots computed from our analytical expression, see Eq.~\ref{E:JTA}.    (d)-(f):  Chronocyclic Wigner function corresponding to panels (a)-(c).  While these plots were computed numerically through Eq.~\ref{E:CWF}, the insets show the corresponding plots computed from our analytical expression, see Eq.~\ref{E:CWFAna}.}
\label{Fig:JTIWgKDP}
\end{figure*}

In Fig.~\ref{Fig:JTIWgKDP} we illustrate the interplay of pump and signal/idler chirp in determining the resulting photon-pair temporal properties, for the particular case of a source based on a KDP crystal and which fulfils asymmetric group velocity matching, similar to that assumed for Fig.~\ref{Fig:JSIJTIKDPBBO}(a) and (b).  We present plots of the joint temporal intensity  $|\tilde{f}(t_i,t_s)|^2$ (panels (a)-(c)),  and of the chronocyclic Wigner function for the heralded single photon (panels (d)-(f)), for a number of different dispersion regimes.  These plots have been computed numerically from Eq.~\ref{E:JSA} and Eq.~\ref{E:CWF}, without resorting to approximations.  The insets show the joint temporal intensity calculated from our analytic expressions (from Eqns.~\ref{E:JTA} and \ref{E:JTAc}, where the plots have been centered at $t_s=t_i=0$).  Note the excellent agreement between the main plots and those in the insets; this tells us that the  approximation used in deriving our analytic expression for the joint temporal amplitude, i.e. Eq.~\ref{E:JTA}, involving expressing the phasemismatch up to first order in frequency detunings, is well justified.  For these plots we have assumed a fixed pump chirp $\beta_p=-4.77\times 10^{-26} s^2$ and a varying degree of signal and idler chirp (for panels (a) and (d), $\beta_s=\beta_i=0$; for panels (b) and (e), $\beta_s=\beta_i=-\beta_p$; for panels (c) and (f), $\beta_s=\beta_i=-2\beta_p$).  Note that the joint spectral intensity $|f(\nu_i,\nu_s)|^2$ exhibits no dependence on external dispersion, i.e. it remains unchanged under the effect of signal/idler dispersion.

\begin{figure}[ht!]
\begin{center}
\includegraphics[width=2.5in]{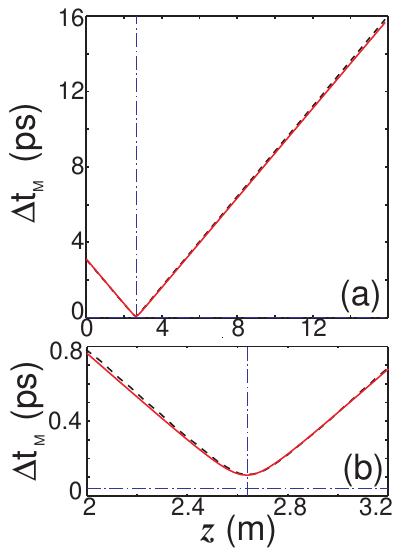}
\end{center}
\caption{(Color online) (a) Temporal duration of heralded single photon in the signal mode, as a function of the chirp parameter which characterizes the propagation of signal and idler photon pairs.  While the dashed black line was computed through a numerical calculation, the red line was computed through our analytic expression (see Eq.~\ref{wtemp}).  In this case the pump is assumed to be chirped with $\beta_p=-4.77 \times 10^{-26}s^2$.  Note that the temporal duration as a function of chirp parameter exhibits a minimum at the propagation distance corresponding to $\beta=-\beta_p$.  The dotted blue line represents the temporal duration for no external dispersion. (b) Close-up of the previous panel in the region of the minimum.}
\label{Fig:duratemp}
\end{figure}

For this source, the dispersion suppression effect described above may be observed.  This effect is clear from panels (a)-(c) and from panels (d)-(f).   Indeed, the temporal duration of the signal photon indicated by panel (b), represents the minimum which can be obtained (for the specific level of pump chirp which has been assumed), and is very close to the temporal duration which would be expected in the absence of pump dispersion.  This can be appreciated from Fig.~\ref{Fig:duratemp}(a), where the temporal duration $\Delta t_M$ of the signal photon is plotted vs the dispersive propagation distance (propagation is assumed to occur in fused silica fiber).   The dotted blue line indicates the temporal duration expected without dispersion. It is clear that there is a specific value of the propagation distance, $z_{min}$, which corresponds to the dispersion matching condition $\beta_s=\beta_i=-\beta_p$ for which the temporal duration reaches its minimum value. In Fig.~\ref{Fig:duratemp}(b) we show a close-up of the plot shown in panel (a), in the region where the minimum occurs.  For this source, asymmetric group velocity matching leads to $\tau_s=0$, while $|\sigma \tau_i| =1.34\times 10^{2}$ so that the condition $|\sigma \tau_i| \gg 1$ may be considered to be fulfilled.  The quantities  $|\sigma^2 \beta_p|$ and $|\sigma \tau_i|$ are of the same order (so that condition $|\sigma^2 \beta_p| \ll |\sigma \tau_i|$ is not fulfilled); therefore, for a smaller pump dispersion, or a longer crystal, the minimum temporal duration can further approach the ideal value obtained without dispersion.

The effect of chirp on the chronocyclic Wigner function may be appreciated from panels (d) and (f) of Fig.~\ref{Fig:JTIWgKDP}.  Specifically, as already discussed in Section~\ref{Sec:Winger},  chirp in general leads to a rotation of this function in chronocyclic space in such a way that the spectral width remains constant, while the temporal width increases.    Note that for the dispersion matching propagation distance, for which $\beta_p=-\beta_s=-\beta_i$, the structure of the CWF and the JTI without dispersion is essentially recovered, despite the presence of external dispersion (i.e. compare Fig.~\ref{Fig:JTIWgKDP}(b) with Fig.~\ref{Fig:JSIJTIKDPBBO}(b), and Fig.~\ref{Fig:JTIWgKDP}(e) with Fig.~\ref{Fig:Wig0D}(a)).

\section{Effects of dispersive propagation on the two-photon state}
\label{Sec:migration}

In this section we explore the effect on the spectral  characteristics of PDC two-photon states of propagation through a dispersive medium.    We will consider the specific case where the signal and idler photons both propagate through the same dispersive medium, such as a fiber, assumed to be non-birefringent so that the two orthogonally-polarized generated photons experience the same dispersion (these results could be easily generalized to differing dispersion for the signal and idler photons, since our treatment in previous sections is valid for both balanced and unbalanced propagation).

The two photon state, after propagation through distance $z$ (e.g. in a fiber) in a medium with quadratic chirp (i.e. ignoring higher order dispersive terms), can be written as

\begin{equation}
|\Psi(z)\rangle=|\mbox{vac}\rangle+\eta\int d \nu_s \int d \nu_i f(\nu_i,\nu_s;z)
|\nu_s\rangle|\nu_i\rangle \label{E:state}
\end{equation}

\noindent in terms of the corresponding JSA

\begin{equation}
f(\nu_i,\nu_s;z)= f(\nu_i,\nu_s;0)
e^{i B z (\nu_i^2+\nu_s^2)}.
\end{equation}

Here, $f(\nu_i,\nu_s;0)$ denotes the JSA for $z=0$ i.e. for no dispersion experienced by the signal and idler photons ($B=0$), as given by Eq.~\ref{E:JSAAn}; $B$ is the GVD parameter defined as $B= \kappa''/2$ where $\kappa$ represents the wavenumber which characterizes the dispersive medium.  Note that $f(\nu_i,\nu_s;0)$ includes any dispersive phases associated with pump chirp.

While the degree of entanglement remains constant during propagation of the signal/idler photons through a lossless medium,  we will show that the degree of modulus-only entanglement, i.e. ignoring any phase entanglement, when calculated in the time domain can vary drastically during propagation.  What this means is that entanglement can be considered to reside to varying degrees in the modulus and in the phase of the joint temporal amplitude depending on the propagation distance $z$.  We refer to this phenomenon as spectral (temporal) entanglement migration, referring to the observation that entanglement can ``migrate'' between the modulus and the phase of the JTA; there is a corresponding effect which occurs in the spatial domain for free-space propagation\cite{chan07}.

In our analysis we first restrict attention to the modulus of the joint amplitude, i.e. we neglect any phase entanglement.
In cases where there is in fact no phase entanglement, then of course the degree of entanglement of the reduced state considered here coincides with the degree of entanglement of the actual physical state.   Also note that the joint spectrum obtained through a phase-insensitive spectrographic measurement such as those reported in Refs.~\cite{kim05,poh07} corresponds to the reduced state considered here.  By carrying out a Schmidt decomposition of the reduced sate and computing the resulting Schmidt number, for different values of the propagation distance $z$, it becomes possible to directly monitor the degree of modulus-only entanglement during propagation of the photon pair.  We refer to the modulus-only Schmidt number calculated in the spectral domain as $K_{m,S}$ and to the modulus-only Schmidt number in the temporal domain as $K_{m,T}$.  Note that while the Schmidt number $K$ calculated in the temporal domain must equal the Schmidt number calculated in the spectral domain, this is no longer true for the modulus-only Schmidt numbers where, in general, $K_{m,T} \neq K_{m,S}$. We also note that while $K_{m,S}$ is constant with respect to the propagation distance $z$ (since the JSI is insensitive to dispersion), in general $K_{m,T}$ is a function of $z$.

While carrying out an analytic Schmidt decomposition on the full two-photon state without resorting to approximations is challenging, this becomes possible for two-photon states for which the JSA can be written in terms of a real-valued Gaussian function of the form $\exp[-(A x^2+B y^2 + C x y)]$ where $x,y$ are variables and $A,B,C$ parameters.  In this case, the
Schmidt eigenfunctions are given by Gauss-Hermite functions
(see Refs.~\cite{uren03, starikov82}), while the eigenvalues are given as  $\lambda_n=(1-\mu^2)\mu^{2n}$ characterized by a parameter $\mu$.  Calculation of the Schmidt decomposition in the temporal domain yields the following expression for this parameter

\begin{equation}
\label{ }
\mu_t=\frac{(\Omega_{i,R}^2 \Omega_{s,R}^2)^{1/2}-(\Omega_{i,R}^2 \Omega_{s,R}^2 -(\Omega_{si,R}^2)^2)^{1/2}}{\Omega_{si,R}^2},
\end{equation}

\noindent in terms of which the reduced Schmidt number (in the temporal domain) is given by

\begin{equation}
K_{m,T}=\frac{1+\mu_t^2}{1-\mu_t^2}.
\label{E:Ksch}
\end{equation}

In order to illustrate this entanglement migration effect, we will consider as specific example the factorable (in the absence of external dispersion) states obtained through asymmetric GVM.

\begin{figure}[ht]
\begin{center}
\includegraphics[width=3.4in]{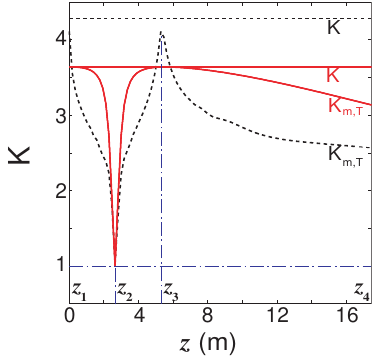}
\end{center}
\caption{(Color online) This plot shows, as a function of the propagation distance $z$,  the Schmidt number $K$ and the reduced temporal Schmidt number $K_{m,T}$; we present both numerical (dashed black line) and analytic (continuous red line) curves for these functions. The minimum allowed value $K=1$ is indicated with a dotted blue line.  While $K$ remains constant during propagation,  $K_{m,T}$ exhibits a rich structure.  From these plots we may infer that entanglement migrates from the modulus of the joint temporal amplitude (at $z=z_1$), to its phase (at $z=z_2$), back to the modulus (at $z=z_3$), and slowly to the phase once more (for large $z$).  }
\label{Fig:KredKDP2}
\end{figure}

Fig.~\ref{Fig:KredKDP2} shows, for a factorable, spectrally elongated source (with the same parameters as in Fig.~\ref{Fig:JSIJTIKDPBBO}(a)),  the dependence of the reduced Schmidt number $K_{m,T}$, calculated through Eq.~\ref{E:Ksch}, on the propagation distance $z$ (through a fused silica fiber). Of course, the dependence of the ``total'' Schmidt number $K$ on propagation distance $z$ is trivial: it remains constant.  Here, we have assumed a value for the quadratic pump dispersion chirp of $\beta_p=-4.77\times10^{-26} s^2$ (note that for vanishing pump chirp, there is no entanglement, and therefore $K_{m,T}$  remains constant at unity).    We observe that there is a specific propagation distance, $z_{min}$,  for which the reduced Schmidt number $K_{m,T}$ essentially reaches the minimum value possible, i.e.  unity, which corresponds to the dispersion matching length for which $\beta_s=\beta_i=-\beta_p$ identified in the previous section.  In the figure we have presented both: i) a plot of $K_{m,T}$ derived from our analytic expression (continuous red line; see Eq.~\ref{E:Ksch}), and ii) a numerical version where we have not resorted to approximations (dashed black line).   For the analytic calculation, we obtain the ``total'' $K$ through Eq.~\ref{E:Visexp}, with $K=1/V$. The plots obtained analytically and numerically exhibit the same general features for $K_{m,T}(z)$: a fast drop from  $z=0$ to $z=z_{min}$ (corresponding to $\beta=-\beta_p$), reaching a value approaching unity, a fast rise to a level similar to that at $z=0$, and a gradual drop for larger propagation distances. Thus, in the temporal domain, as the photon pair propagates through the fiber, entanglement migrates from the modulus to the phase, back to the modulus, and finally gradually to the phase once more.  This behavior is, of course, also apparent from Fig.~\ref{Fig:JTIWgKDP} panels (a)-(c), where the first and last panels show strong temporal (modulus) correlations, while the dispersion-matched case shows essentially no temporal (modulus) correlations.  Note that a similar behavior to that observed in Fig.~\ref{Fig:KredKDP2} for $K_{m,T}$ would be observed in terms of the Fedorov ratio $\mathcal{F}$ (see Eq.~\ref{E:Fedorov}).

Let us note that for the dispersion-matching propagation distance in the fiber, $z=z_{min}$, $K_{m,T} \approx 1$ implies that in the time domain entanglement resides entirely in the phase.  Furthermore, as discussed above for this particular source,  when viewed in the spectral domain, entanglement resides entirely in the phase for any propagation distance $z$.  Thus, remarkably, for this particular source at $z=z_{min}$ entanglement resides entirely in the phase in \textit{both} domains.  We speculate that applications of this form of phase-only entanglement, in both the spectral and temporal domains, may result from the fact that it is less accessible to measurements than more typical modulus entanglement.


\section{Conclusions}

In this paper we have presented a study of the effects of external dispersion, i.e. introduced by optical elements other  than the nonlinear medium itself, for the process of type II, collinear spontaneous parametric downconversion.    We have derived expressions for the joint amplitude of the photon pair state, both in the spectral and temporal domains.   In particular, we present an expression for the joint temporal amplitude resorting to a first-order expansion of the phasemismatch, as well as a corresponding expression where we also resort to the Gaussian approximation for the phasematching function.   While the first one yields two-photon characteristics remarkably similar to those obtained through a numerical calculation which does not resort to approximations, the second one permits analytic expressions for the following:  i)  Hong-Ou-Mandel interferogram, involving two heralded single photon sources, ii) Chronocyclic Wigner function for heralded single photons, which fully characterizes the spectral (temporal) properties of the single photons and iii) Schmidt decomposition of the modulus of the joint temporal amplitude.
Exploiting this description, we find that pump chirp can represent an extremely useful tool for controlling the degree of photon-pair entanglement.  Indeed, we find that if the two-photon state is factorable in the absence of pump chirp, then the resulting Schmidt number (which quantifies the entanglement present) can be continuously adjusted, in principle, from $K=1$ to any desired value through the pump chirp parameter.  We likewise predict a dispersion suppression effect which occurs for two-photon states designed to be factorable and spectrally-elongated: the temporal duration of the heralded single photon in the presence of pump chirp can approach the temporal duration which would have been observed without pump chirp, if a specific condition is fulfilled by the pump and PDC chirp parameters.

We explicitly show the relationship between: i) the photon pair Schmidt number, ii) the Hong-Ou-Mandel visibility in a two-crystal arrangement and iii) the heralded single photon purity.   This gives us a method to experimentally characterize both the degree of photon-pair entanglement and the degree of heralded single photon purity.   We show that the spectral (temporal) entanglement present in a two photon state can migrate between the modulus and the phase of the joint temporal amplitude. We present details of a specific source for which entanglement is expected to reside entirely in the phase in both the spectral and temporal domains.    We believe that the analysis presented in this paper will be useful for the design and implementation of specific photon-pair sources tailored for quantum information processing applications.

\begin{acknowledgements}
We acknowledge support from Conacyt grant 46370-F and UC-MEXUS grant CN-06-82.
\end{acknowledgements}

\appendix

\section{}

The analytic expression for the joint amplitude (see Eq.~\ref{E:JTA}) may be written in terms of a function of the form

\begin{eqnarray}
&Z(G,x_0;x)=\frac{1}{\sqrt{\pi}} \int\limits_{G(x-x_0)}^{G(x+x_0)} dt \ e^{-t^2} \nonumber \\
&= \frac{1}{2}\left\{\mbox{erf}[G(x+x_0)]-\mbox{erf}[G(x-x_0)] \right \}
\end{eqnarray}

If we write parameter $G$ in terms of its modulus and phase, i.e. $G=|G|\exp(i \theta)$, this function may be shown to converge for $-\pi/4 \le \theta \le \pi/4$.

In the limit where $|G x_0|\to\infty$, function $Z(G,x_0;x)$ behaves as a ``top hat'' function

\begin{equation}
\lim_{|G x_0|\to\infty} |Z(G,x_0;x)|=\mbox{rect}(-x_0,x_0;x)
\end{equation}

where
\begin{equation}
\mbox{rect}(x;a,b) = \left\{
\begin{array}{c l}
  1, \mbox{  } a \le x \le b\\
  0, \mbox{   otherwise}
\end{array}
\right.
\end{equation}

In the limit where $|G x_0|\to\ 0$, function $Z(G,x_0;x)$ behaves as a Gaussian function.

\begin{equation}
\lim_{|G x_0| \to 0} Z(G,x_0;x)=\frac{2}{\sqrt{\pi}} G x_0 \exp(-G^2 x^2)
\end{equation}

\end{document}